\documentclass[twocolumn,pre,showpacs,preprintnumbers]{revtex4-1}

\usepackage{graphicx}
\usepackage{dcolumn}
\usepackage{bm}
\usepackage{amsmath}  
\usepackage[abs]{overpic}
\usepackage{aas_macros}
\usepackage{soul}

\begin{document}

\bibliographystyle{plainnat}

\title{Spectrum of kinetic plasma turbulence at  0.3$-$0.9 astronomical units from the Sun} 

\author{Olga Alexandrova\textsuperscript{1}, 
Vamsee Krishna Jagarlamudi\textsuperscript{1,2,3}, 
Petr Hellinger\textsuperscript{4,5},
Milan Maksimovic\textsuperscript{1}, 
Yuri Shprits\textsuperscript{6} \& 
Andre Mangeney\textsuperscript{1}}
\affiliation{\textsuperscript{1}LESIA, Observatoire de Paris, Universit\'e PSL, CNRS, Sorbonne Universit\'e, Universit\'e de Paris, 5 place Jules Janssen, 92195 Meudon, France.}
\affiliation{\textsuperscript{2}LPC2E, CNRS, University of Orl\'eans, 3 Avenue de la Recherche Scientifique, 45071 Orleans Cedex 2, France.}
\affiliation{\textsuperscript{3}National Institute for Astrophysics-Institute for Space Astrophysics and Planetology, Via del Fosso del Cavaliere 100, I-00133 Roma, Italy}
\affiliation{\textsuperscript{4}Astronomical Institute, CAS, Bocni II/1401, Prague CZ-14100, Czech Republic}
\affiliation{\textsuperscript{5}Institute of Atmospheric Physics, CAS, Bocni II/1401, Prague CZ-14100, Czech Republic}
\affiliation{\textsuperscript{6}GFZ German Research Centre for Geosciences, University of Potsdam, D-14469 Potsdam, Germany.}

\date{2 June 2021}

\begin{abstract}

We investigate  spectral properties of turbulence in the solar wind that is a weakly collisional astrophysical plasma, accessible to \emph{in-situ} observations. 
Using the Helios search coil magnetometer measurements in the fast solar wind, in the inner heliosphere, we focus on properties of the turbulent magnetic fluctuations at scales smaller than the ion characteristic scales, the so-called \emph{kinetic plasma turbulence}. 
At such small scales, we show that the magnetic power spectra between 0.3 and 0.9~AU from the Sun have a generic shape $\sim f^{-8/3}\exp{(-f/f_d)}$ where the dissipation frequency $f_d$ is correlated with the Doppler shifted frequency $f_{\rho e}$ of the electron Larmor radius. This behavior is statistically significant: all the observed kinetic spectra are well described by this model, with $f_d = f_{\rho e}/1.8$.  Our results indicate that the electron gyroradius plays	the role of the dissipation scale and marks	the end of the electromagnetic	cascade in the solar wind.


\end{abstract}
\maketitle

\section{Introduction}
Astrophysical plasmas are often very rarefied so
that the Coulomb collisions are infrequent \cite[e.g.,][]{meyer-vernet07,Salem2003}:
in contrast to the usual neutral fluids,
the collisional dissipation (viscous and resistive) channels
are weak, and the Kolomogorov’s dissipation scale \cite{frisch1995} is ill-defined.
Furthermore, the presence of a background magnetic field $\bm{B_0}$ introduces
a preferred direction \cite[e.g.,][]{shebalin83,oughton94,Verdini2015,Oughton_2020}
and allows the existence of propagating incompressible modes (Alfv\'en waves). 
The different plasma ion and electron constituents have a number of characteristic (kinetic) scales at which properties of turbulent fluctuations change.

Considering all this complexity, one may wonder whether there is a certain degree of generality in space plasma turbulence. 
In particular, does the dissipation range have a general spectrum, 
as  is the case in  neutral fluid turbulence \cite{chen93,frisch1995}?

The solar wind plasma, which is accessible to \emph{in-situ} space exploration, 
has proven to be a  very useful laboratory to study the  astrophysical plasma turbulence \cite[e.g.,][]{bruno13,alexandrova13}. 
Since the first early in-situ measurements, \cite[e.g.,][]{Coleman1968}, our knowledge of the large-scale turbulence in the solar wind has greatly improved, \cite[e.g.,][]{bruno13,kiyani15}. There is an extended inertial range of scales 
at which  incompressible magnetohydrodynamics (MHD) phenomenologies \cite{goldreich95,boldyrev05,chandran15}, similar in spirit to  Kolomogorov's phenomenology, may be invoked to understand the formation  of a Kolmogorov-like spectrum of magnetic fluctuations $\sim k^{-5/3}$. 
(Note that satellite measurements are time series, thus, in Fourier space one gets frequency spectra. At the radial distances from the Sun studied here,  any characteristic plasma velocity, except whistler wave phase speed, is less than the solar wind speed $V$. Thus, one can invoke  Taylor’s hypothesis and convert a spacecraft-frame frequency $f$ to a flow-parallel wavenumber $k$ in the plasma frame $k=2\pi f/V$.) 

	At the short wavelength end of the inertial domain, i.e., at scales of the order of the proton inertial scale $\lambda_p=c/\omega_{pp}$ (where $c$ is the speed of light  and $\omega_{pp}$ is the proton plasma frequency) the spectrum steepens. 
At these scales ($\sim 100$~km at 1~AU from the Sun \footnote{One astronomical unit (AU) is the distance from Earth to the Sun, which is about $1.5 \times 10^{11}$~m.}), the MHD approximation is no longer valid; the ``heavy'' ion (basically, a proton in the solar wind)  fluid and the ``light'' electron fluid behave separately,  \cite[e.g.,][]{Matthaeus2008PRL,Hellinger2018ApJL,Papini2019ApJ}.  
It is still not completely clear whether the spectral steepening at ion scales is the beginning of the dissipation range or a transition to another cascade taking place between ion and electron scales or a combination of both \cite[e.g.,][]{alexandrova13,chen16_rev,Li-Howes2019-JPP}. 
Recent von K\'arm\'an-Howarth analyses of direct numerical simulations and \textit{in-situ} observations \cite{Hellinger2018ApJL,Bandyopadhyay2020PRL} indicated  that the transition from the MHD inertial range to the sub-ion range is due to a combination of the onset of the Hall MHD effect and  a reduction of the cascade rate likely due to some dissipation mechanism. 
Then, the question arises as to how much of the dissipation of the turbulent energy is flowing into the ions and how much is flowing into the electrons. 
In the vicinity of the electron scales ($\sim 1$~km at 1~AU), the fluid description 
no longer holds,  and the electrons should be considered as particles. The present paper focuses on this short wavelength range, i.e., between the ion scales and a fraction of the electron scales.

The first solar wind observations of turbulence at scales smaller than ion scales (the so-called \textit{sub-ion scales})
 were reported by \citet{denskat83}, using the search coil magnetometer (SCM) on Helios space mission at radial distances $R\in[0.3,0.9]$~AU from the Sun. 
From this pioneering work we know that  between the ion and electron scales,  the magnetic spectrum follows  an $\sim f^{-3}$ power law. 

Thanks to the Spatio-Temporal Analysis of Field Fluctuations (STAFF)  instrument on Cluster space mission  \citep{escoubet97,cornilleau-wehrlin97}, 
which is the most sensitive SCM flown in the solar wind to date, the small scale tail of the electromagnetic cascade at 1~AU could be explored  down to  a fraction of electron scales $\sim 0.2 - 1$~km 
\citep{mangeney06,alexandrova08_angeo,alexandrova09,alexandrova12,alexandrova13, sahraoui10,sahraoui13,lacombe17,Matteini2017}, i.e., up to 1/5 of electron scales.
These observations  seem confusing at first glance: 
the spectral shape of the magnetic fluctuations varies from 
one record  to another,  suggesting 
that the spectrum is not universal at kinetic scales \citep{mangeney06,sahraoui10,sahraoui13}. 
However, as was shown in \citep{lacombe14,roberts17,Matteini2017},  
most of these spectral variations are due to the presence, or absence, of quasi-linear whistler waves  
with frequencies at a fraction of the electron cyclotron frequency 
	$f_{ce}=e B_0/(2\pi m_e)$ (where $e$ and $m_e$ are the charge and the mass of an electron, respectively)
and wave vectors $\bm{k}$ quasi-parallel to $\bm{B_0}$ \citep{lacombe14}. 
These waves may result from the development of some instabilities associated with either an increase of  the electron temperature anisotropy or an increase of the electron heat flux in some regions of the solar wind  \cite{Stverak2008JGR}.
In the absence of whistlers, the background turbulence is characterized by  low frequencies in the plasma frame  and wave vectors mostly perpendicular to the mean field $\bm{k} \perp \bm{B_0}$~\citep{lacombe17}. 
This quasi-2D  turbulence is convected by the solar wind  (with the speed $\bm{V}$) 
	across the spacecraft and appears in the satellite frame at frequencies $f=k_{\perp} V/2\pi$. 
 It happens that these frequencies are below but close to $f_{ce}$, exactly in the range where whistler waves (with $\bm{k} \; \| \; \bm{B_0}$ and $f\simeq (0.1-0.2) f_{ce}$) may appear locally. 
 Therefore, the superposition of turbulence and whistlers at the same frequencies is coincidental. 
 If we could perform measurements directly in the plasma frame, these two phenomena  would be completely separated in $\bm{k}$ and $f$. 
 A possible interaction between turbulence and whistlers is out of the scope of the present paper. 
 We focus here on the background turbulence at kinetic scales only.

A statistical study by \citet{alexandrova12} of solar wind streams at 1~AU under different plasma conditions  showed that, in the absence of parallel  whistler waves,
the quasi-2D background turbulence forms a spectrum 
$\sim k_{\perp}^{-8/3}\exp{(-k_{\perp}\ell_d)}$, 
with a cut-off scale $\ell_d$ well correlated with the electron Larmor radius 
 $\rho_e = \sqrt{2k_B T_{e\perp}/m_e}/(2\pi f_{ce})$ (where $k_B$ is the Boltzmann constant and $T_{e\perp}$ is the electron perpendicular temperature).
Such a spectrum with an exponential correction indicates 
a lack of spectral self-similarity
at electron scales, as in the dissipation range of the neutral flow turbulence.  
How general  is this kinetic spectrum? 
Is it observed closer to the Sun than 1~AU?

Parker Solar Probe (PSP) observations in the slow wind at $0.17$~AU show a spectrum close to $\sim f^{-8/3}$ at sub-ion scales \cite{Bale2019Nat}. 
In a  statistical study of turbulent spectra up to 100~Hz,  \citet{Bowen2020PRL}  
determined spectral indices up to 30~Hz, confirming a power law usually observed at 1~AU $\sim f^{-2.8}$ 
\cite{alexandrova09,chen10a,alexandrova12,kiyani13,sahraoui13}.  
The PSP--SCM data products up to 100~Hz used in \citep{Bale2019Nat,Bowen2020PRL} 
and the instrumental noise level  do not allow the resolution of electron scales at $0.17$~AU,
at  least for the types of solar wind and the  Sun-spacecraft distances sampled by PSP to date.

In this paper, we analyze magnetic spectra  within the $[7,700]$~Hz range at radial distances between 0.3 and 0.9~AU thanks to Helios measurements.  
Here, for the first time, we provide a turbulent spectrum at electron scales and its simple empirical description at distances from the Sun smaller than 1 AU. 
The spectrum follows a function similar to that found at 1~AU, indicating generality of the phenomenon.

\section{Data}
The SCM instrument on Helios space mission \cite{Neubauer1977-zg} 
consists of three orthogonally oriented search coil sensors which  are mounted on a boom at a distance of 4.6~m from the center of the spacecraft with the $z$-sensor parallel to the spin axis and $x$ and $y$ sensors in the spin plane. The wave forms from the sensors are processed in an on-board spectrum analyzer. They pass through 8 band-pass filters which are continuous in frequency coverage and logarithmically spaced. The central frequencies of the 8 channels are 6.8, 14.7, 31.6, 68, 147, 316, 681 and 1470~Hz. The novel feature for the time of construction of the instrument was that the filter outputs were processed by a digital mean-value-computer on board of Helios 
	 \cite{Neubauer1977-jgr}.

Thus, the instrument provides magnetic spectra for two of three components, $(B_y, B_z)$ and rarely $(B_x, B_z)$, in the Spacecraft Solar Ecliptic reference frame, which is equivalent to the Geocentric Solar Ecliptic frame \footnote{The Geocentric Solar Ecliptic (GSE) frame has its $x$-axis pointing from the Earth towards the Sun,  the $y$-axis is chosen to be in the ecliptic plane pointing towards dusk (thus opposing planetary motion), $z$-axis is normal to the ecliptic plane, northwards.}.   The available Helios-SCM products are the spectra integrated over 8~s.
For the present study we use only the spectra of $B_y$. Indeed, the pre-flight noise level for the $B_y$ spectra matches well  the post-flight noise level, which is not the case for $B_z$. More details on the instrument and data processing can be found in \citep{Neubauer1977-jgr}.

We have analyzed  246543 individual $B_y$--magnetic spectra as measured by SCM on  Helios--1
with signal-to-noise ratios (SNR) larger than or equal to 2 up to 100 Hz, at radial distances from the Sun $R\in[0.3,0.9]$~AU 
Among them,  about $ 2\%$  of the spectra show spectral bumps between the lower hybrid frequency $f_{lh}$ and  $\sim 0.25 f_{ce}$ \cite{Jagarlamudi2020ApJ}. 
Such bumps are the signatures of parallel whistler waves as was shown in \cite{lacombe14}. 
The analysis of these spectra with bumps, shows that the signatures of whistlers are mostly present in the slow wind ($V<500$~km/s) and their appearance  increases with the distance from the Sun \cite{Jagarlamudi2020ApJ}. 
In the fast wind ($V>600$~km/s) and close to the Sun, we do not observe signatures of whistlers in 8-s individual spectra of Helios--SCM.  
Here, we analyze background turbulence spectra in the fast solar wind, i.e.,  without signatures of whistler waves.  

On the basis of this first analysis of 246543 $B_y$--spectra with a SNR~$\ge 2$ up to 100~Hz, we can already say that the background turbulence without signatures of whistlers is commonly observed (98\% of the analyzed spectra) and its spectral shape is very similar at different radial distances as we will see below, just the amplitude changes. 
Turbulent level decreases with radial distance \cite{beinroth81,denskat83,bourouaine12,Chen_2020ApJS}
and thus further from the Sun, fewer SCM frequencies are resolved. 
For the statistical study, we will consider 3344 spectra with a SNR larger than or equal to 3 up to $316$~Hz and among them 39 spectra with a  SNR~$\ge 3$ up to $ 681 $~Hz. All these 3344 spectra are at 0.3~AU.

\section{Spectral analysis}

\begin{figure}
	\includegraphics[width=8.6cm]{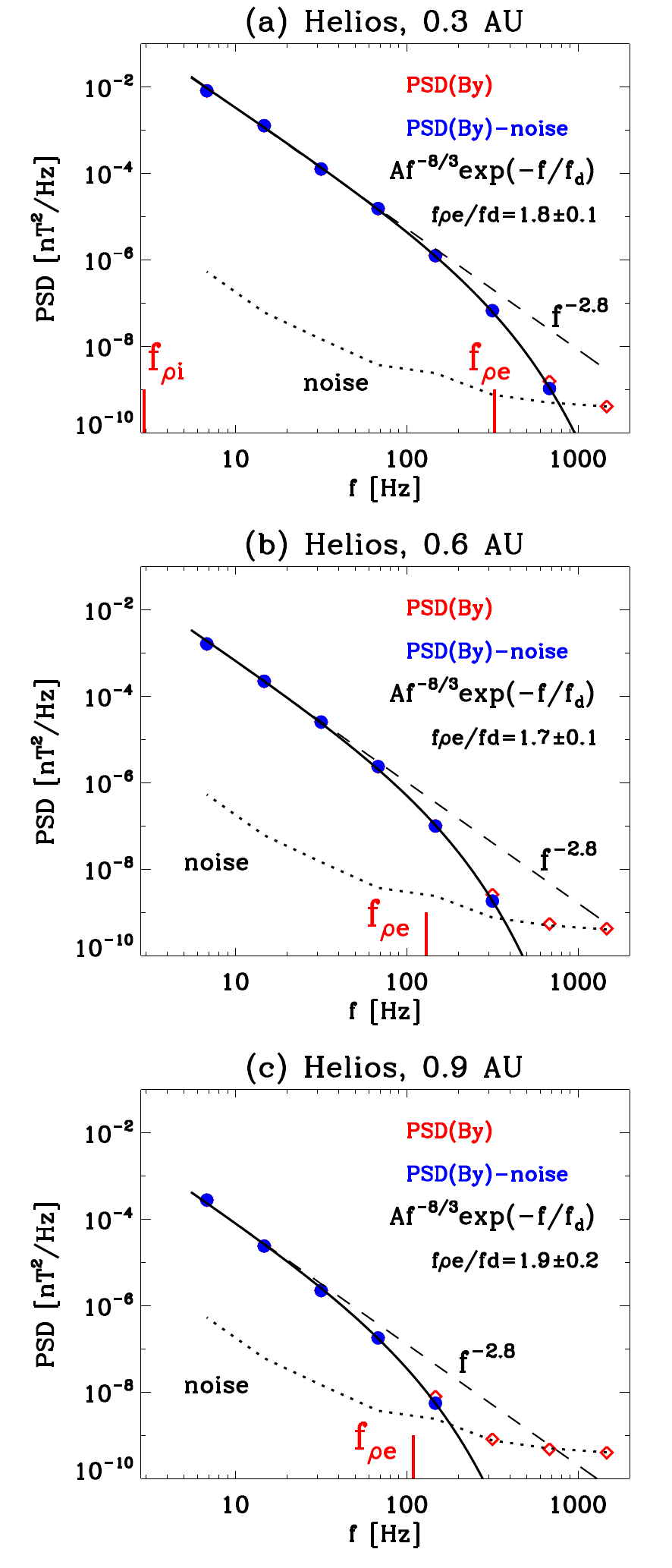}
	\caption{Examples of the most intense Helios--SCM spectra of $B_y$ component, as functions of the spacecraft-frame frequency $f$, at (a) 0.3~AU, (b) 0.6~AU and (c) 0.9~AU. For the 3 radial distances, the raw-spectrum is shown by red diamonds, the corrected spectrum, after the subtraction of the noise -- by blue dots, the black solid line gives the fit with the model function~(\ref{eq:model}),  the dashed line gives $f^{-2.8}$ power-law  for comparison and the dotted line indicates the noise level of the Helios-SCM-$B_y$. 
		Vertical red lines give the Doppler shifted kinetic scales: in (a), $\rho_p$ and $\rho_e$ appear at $f_{\rho p} = 2.9$~Hz and $f_{\rho e}= 325$~Hz, respectively;  in (b) they appear at  $f_{\rho p}\simeq 1$~Hz  and $f_{\rho_e}=130$~Hz, respectively; and in (c) they appear at $f_{\rho p}\simeq 1$~Hz and $f_{\rho_e}=110$~Hz, respectively.}
	\label{fig:helios-examples}
\end{figure}

Figure~\ref{fig:helios-examples}(a)-(c) show examples of the most intense $B_y$--spectra as measured by SCM on Helios--1 at 0.3, 0.6 and 0.9~AU, respectively. 
For the 3 radial distances from the Sun the raw power spectral densities (PSDs) are  shown by red diamonds. The dotted line indicates the noise level of the instrument for the $B_y$--component.  
The spectra corrected for the noise contribution by the subtraction of the noise level are shown by blue dots. 
Vertical red lines give the Doppler shifted kinetic scales. Plasma parameters, characteristic lengths and frequencies corresponding to these spectra 
	 are given in Table~\ref{tab:tab-fig1}. 

We perform a least square fit of the 3 corrected spectra with 
the model function known to describe the kinetic spectrum at 1~AU \cite{alexandrova12}:
\begin{equation}\label{eq:model}
P_{\text{model}}(f) = Af^{-8/3}\exp{(-f/f_d)}.
\end{equation}
This model has two free parameters: the amplitude of the spectrum $A$ and 
the  dissipation frequency $f_d$.  
The result of this fitting is shown by a black solid line in the 3 cases. The corresponding maximal physical frequencies $f_{max}$ (the highest frequency where the SNR is $\ge 3$ still verifies \cite{alexandrova10}) together with the results of the fit are given  
at the end of Table~\ref{tab:tab-fig1}. 
At 0.3~AU, the spectrum is well resolved up to $f_{max}=681$~Hz (the 7th out of the 8 SCM frequencies). The electron Larmor radius $\rho_e\simeq 0.4$~km appears at $f_{\rho e} = V/(2\pi\rho_e) = 325$~Hz (see the right vertical red  line). 
	Thus, in this case, turbulence is resolved up to a minimal  scale of about 
$\ell_{min} = V/(2\pi f_{max}) = 0.47\rho_e$ (see the bottom row of Table~\ref{tab:tab-fig1}). 
As expected \cite{beinroth81,denskat83,bourouaine12,Chen_2020ApJS}, 
further from the Sun the intensity of the spectra decreases with $R$: at 0.6~AU, the spectrum is resolved up to 316~Hz and at 0.9~AU, it is resolved only up to 147~Hz. 
In both cases, nonetheless, the electron Larmor radius is resolved  as $\rho_e\sim 1/B_0$ 
increases with $R$ and the corresponding frequency $f_{\rho e}$ decreases (see vertical red lines in  Figure~\ref{fig:helios-examples}(b) and (c): $f_{\rho e} = 130$~Hz at 0.6~AU and $110$~Hz at 0.9~AU).
The observed spectra at 3 radial distances from the Sun are well described by the model,   
and the dissipation frequency $f_d$ decreases from $(183\pm 5)$~Hz at 0.3~AU 
to $(56\pm 4)$~Hz at $0.9$~AU,  following $f_{\rho e}$. 

 From Table~\ref{tab:tab-fig1} one can see that  further from the Sun, the relative errors on free parameters of the fit, $\Delta f_d/f_d$ and $\Delta A/A$, increase, while the $f_{max}$ decreases. This error increase  is expectable: $f_{max}$ is proportional to the turbulence level, and the lower turbulence level corresponds to the smaller SNR and automatically to a smaller number of frequencies to fit; thus, we get higher errors.

\begin{table}
	\centering
	\caption{Plasma parameters, characteristic scales and frequencies, maximal resolved frequency by Helios/SCM, $f_{max}$, and results of the fit to Eq.(\ref{eq:model}) at 3 radial distances from the Sun, corresponding to the spectra in Figure~\ref{fig:helios-examples}. The two bottom rows indicate a fraction of $\ell_d$ and $\rho_e$--scales resolved by these spectra.}
	\begin{tabular}{ c | c c c c c c c c  }
		\hline
								& & & \\
			$R$ (AU)		& \;\;  0.9  \;\; & \;\;  0.6 \;\;  & \;\;  0.3  \;\;   \\ 
									& & & \\ \hline
		$B_0$ (nT) & $8.5$ & 11.6 & $32.2$   \\ 
		$V$ (km/s)  & $720$ & 710 & $740$  \\ 
		$n_p$ (cm$^{-3}$) & $4.8$ & 7.0 & $28.4$  \\ 
		$T_p$ (eV) &$34.3$ & 51.1&$61.2$  \\ 
		$T_e$ (eV) &$9.3$ & 12.7&$12.9$  \\ 
		$T_{p \perp}$ (eV) &$41.2$ & 67.8  &$80.3$  \\ 
		$T_{e \perp}$ (eV) &$7.0$ & 9.0&$12$ \\ 
		$\beta_{p,\perp}$ & $1.1$ & 1.4& $0.9$  \\ 
		$\beta_{e,\perp}$  &$0.2$ &0.2 &$0.13$ \\ 
		$\lambda_p$ (km) &$99$ & 82&$41$   \\ 
		$\rho_p$ (km) &$109$ &102 &$40$ \\ 
		$\lambda_e$ (km)  &$2.3$ & 1.9& $1$  \\ 
		$ \rho_e$ (km) &$1.0$ & 0.9&$0.4$   \\ 
		$f_{cp}$ (Hz) &$0.10 $ &0.2 &$0.5$   \\ 
		$f_{\lambda p}$ (Hz)  &$1.2 $ & 1.4& $2.9 $ \\ 
		$f_{\rho p}$ (Hz) &$1.0$ & 1.1&$2.9$   \\ 
		$f_{\lambda e}$ (Hz)  & $50$& 59&$124$   \\ 
		$f_{\rho e}$ (Hz) &$110$ &130 &$325$  \\ 
		$f_{ce}$ (Hz) &$238$&325 & $900$   \\ 
								& & & \\
		$f_{max}$ (Hz) &147&316 & 681   \\ 
		$A$ (nT$^2/$Hz)Hz$^{8/3}$ &$0.04$&$0.34$ & $1.63$   \\ 
		$\Delta A/A$ &2&0.2 & 0.03   \\ 
		$f_{d}$ (Hz) &$56$&$58$ & $183$   \\ 
		$\Delta f_{d}/f_d$ &0.07&0.04 & 0.03   \\ 
		$f_{d}/f_{max}$ &0.38&0.27 & 0.27   \\ 
		$f_{\rho e}/f_{max}$ &0.74&0.40 & 0.47   \\ \hline
	\end{tabular}
	\label{tab:tab-fig1}
\end{table}

Now let us consider the most intense spectra, i.e., with a SNR that is~$\ge 3$ up to $681$~Hz and with simultaneous measurements of $\bm{B_0}$.  These conditions are verified for 39 spectra at 0.3~AU in the fast wind,  measured during the closest approach of Helios to the Sun. 

\begin{figure}
	\includegraphics[width=8.6cm]{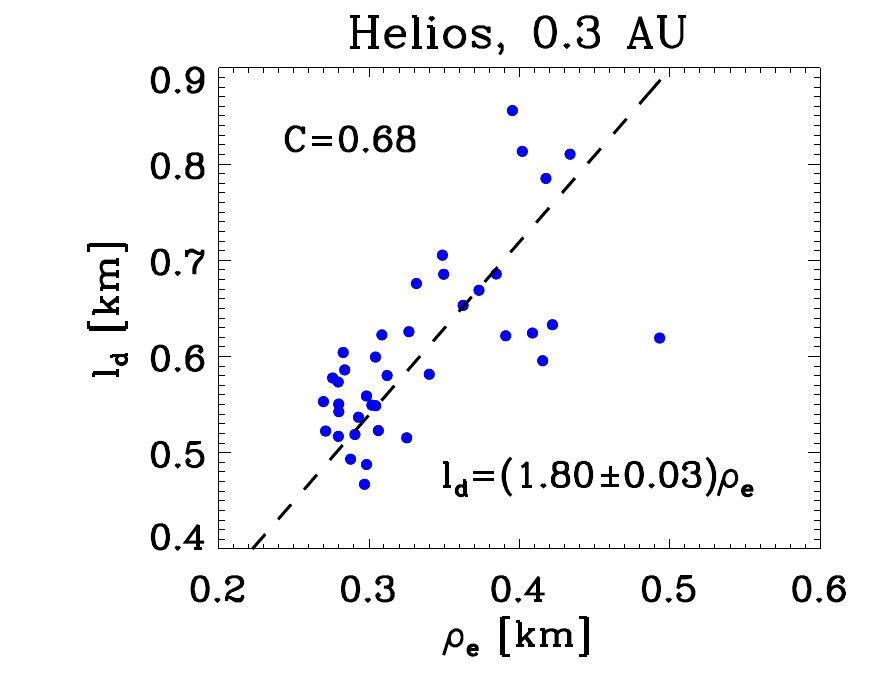} 
	\caption{Results of the fitting procedure of the most intense spectra at 0.3~AU with Eq.~(\ref{eq:model}): dissipation scale $\ell_d = V/2\pi f_d$ as a function of the electron Larmor radius $\rho_e$; 
		the linear dependence $\ell_d = 1.8 \rho_e$ is indicated by the dashed line, with the correlation coefficient $C=0.68$. 
	}
	\label{fig:corr}
\end{figure}


All these spectra are similar to that shown in Figure~\ref{fig:helios-examples}(a). 
We perform a least squares fit of the 39 spectra with the model function,  Eq.~(\ref{eq:model}). The relative errors, $\Delta f_d/f_d$ and $\Delta A/A$, vary between 0.01 and 0.14. 
The dissipation scale $\ell_d$ can be estimated using the Taylor hypothesis $\ell_d = V/(2\pi f_d)$.  
It is found to be correlated with the $\rho_e$ scale  with a correlation coefficient $C =0.68$.   
The relation  $\ell_d \sim 1.8 \rho_e$  is observed  (see Figure~\ref{fig:corr}). 
There is no correlation with the electron inertial length $\lambda_e$ ($C= 0.02$, not shown). 
Thus, we can fix $f_d$ in Eq.~(\ref{eq:model}):
 \begin{equation}\label{eq:model-1param}
P_{\text{model}}(f)=Af^{-8/3}\exp{(-1.8 f/f_{\rho e})}.
\end{equation}
Let us now verify whether this simpler model 
describes a larger statistical sample.

\begin{figure}
	\begin{center}
		\includegraphics[width=9cm]{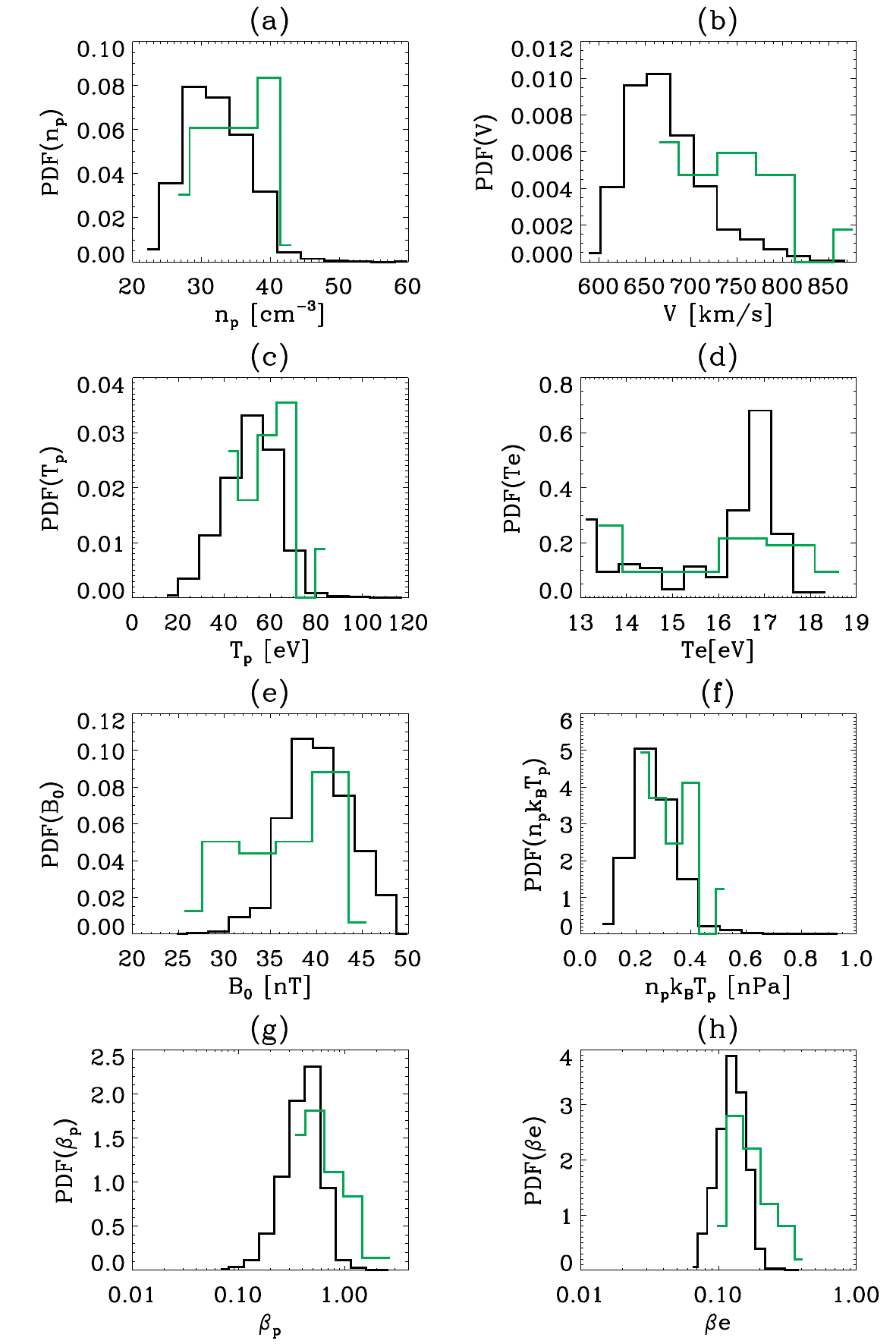}
		\caption{
			Probability distribution functions (PDF's) of the mean plasma parameters at 0.3~AU for the 
			3344 spectra shown in Figure~\ref{fig:spec3344} (black lines) 
			and for the 39 most intense spectra (green lines): (a) proton density $n_p$, 
			(b) solar wind speed $V$, (c) proton temperature $T_p$, (d) electron temperature $T_e$, 
			(e) magnetic field magnitude $B_0$, (f) proton thermal pressure $n_pk_BT_p$, 
			(g) proton plasma beta $\beta_p$, (h) electron beta $\beta_e$.
		}
		\label{fig:spec3344-plasma}
	\end{center}
\end{figure}

To increase the number of spectra analysed, we now also consider less resolved spectra, i.e., with a signal-to-noise ratio larger than 3 up to 316~Hz, 
and with plasma measurements in the vicinity of the spectra 
(i.e., the mean field at most within 16~s around the measured SCM spectrum,  the 
electron temperature $T_e$ within about 30~min; and when not available, 
$T_e$ is taken within a longer time interval but within the same wind type). 
These conditions are verified for 3344 spectra at 0.3~AU in the fast wind.  
Probability distribution functions (PDFs) of the  mean plasma parameters for the 3344 spectra are shown in Figure~\ref{fig:spec3344-plasma} with black lines and those for the 39 most intense spectra analyzed above, are shown by green lines. The proton $\beta_p$  (electron $\beta_e$) plasma beta is the ratio between the proton (electron) thermal pressure and the magnetic pressure. 
From these PDFs, we see that  the 39 most intense spectra are observed for the solar wind with $V> 650$~km/s, for the proton thermal pressure $n_pk_BT_p \ge 0.2$~nPa and for the largest $\beta_p$ and $\beta_e$ values of the analyzed data set (for $\beta_p \ge 0.3$ and $\beta_e \ge 0.1$).

\begin{figure}
	\begin{center}
		\includegraphics[width=8.6cm]{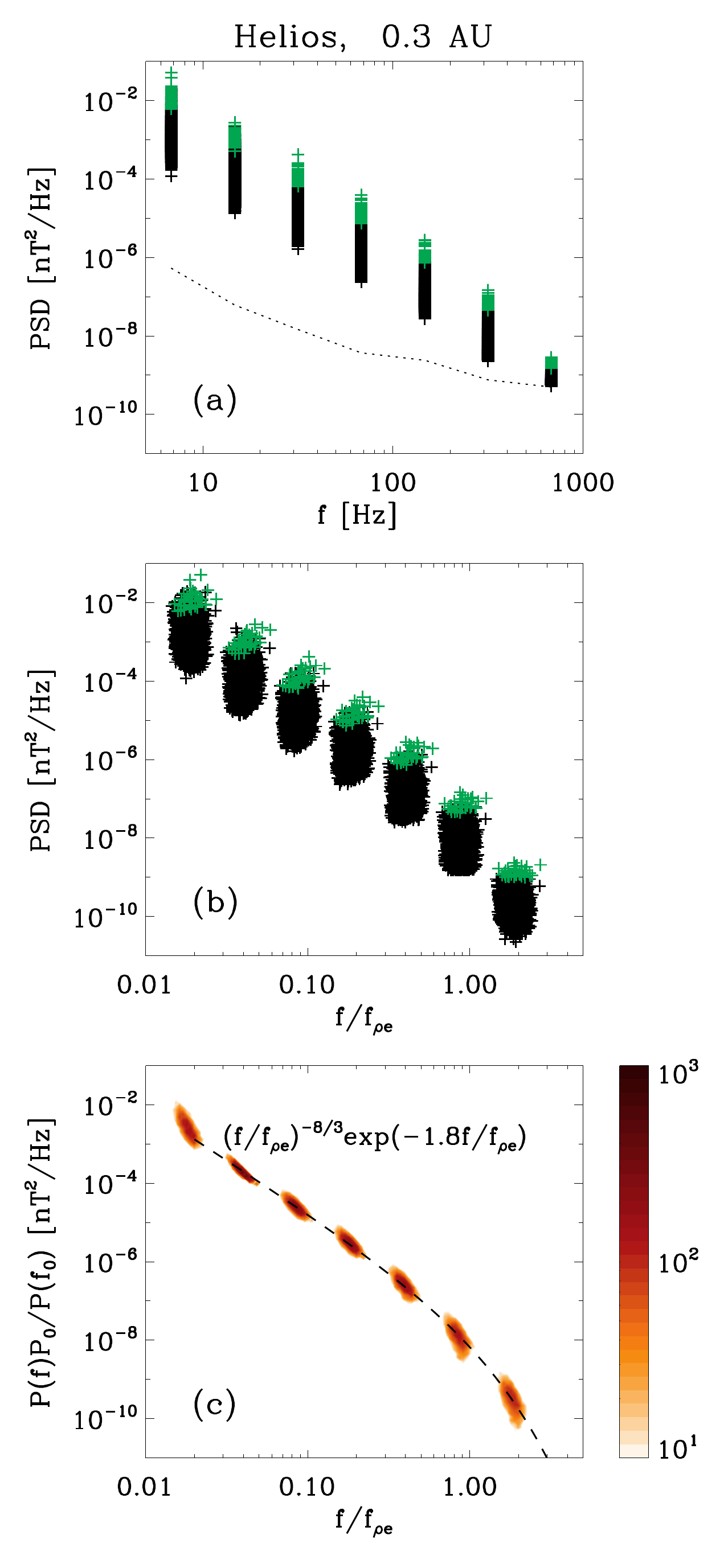}
		\caption{(a) 3344 individual Helios--1 SCM spectra of $B_y$
			as functions of the spacecraft-frame frequency $f$
			at 0.3 AU in the fast wind;  the 39 most intense spectra  are marked by green crosses; the SCM noise for $B_y$ component  is indicated by a dotted line. 
			(b) These 3344 spectra corrected for the noise contribution as functions of $f$ normalised to the Doppler shifted electron Larmor radius
				frequency $f_{\rho e}=V/(2\pi \rho_e)$.
			(c) The same spectra, rescaled by their amplitude at $f_0=0.051 f/f_{\rho e}$ (see the text);  the result is shown as a 2D histogram with the number of the data points proportional to the darkness of the red colour. 
	The dashed line displays the model function, Eq.(\ref{eq:3}).}
		\label{fig:spec3344}
	\end{center}
\end{figure}


Figure~\ref{fig:spec3344}(a) displays the 3344 raw $B_y$ spectra, $P_{\text{raw}}(f)$, 
by crosses. 
	The 39 most intense spectra 
	are marked by green crosses; the noise level for $B_y$, $P_{\text{noise}}(f)$ is indicated by the dotted line.
Figure~\ref{fig:spec3344}(b) shows these 3344 spectra 
corrected for the noise contribution, $P(f)=P_{\text{raw}}(f)-P_{\text{noise}}(f)$, and as functions of $f$ normalised
to the Doppler shifted electron Larmor radius frequency,
$f_{\rho e}=V/(2\pi \rho_e)$. Let us now superpose all spectra together. 
Figure~\ref{fig:spec3344}(c) shows a 2D histogram calculated with the spectra of the middle panel and rescaled by their amplitude at $f/f_{\rho e}=0.051$, i.e., $P(f)P_0/P(f_0)$. This means that by construction all spectra pass through the point $(f_0,P_0)=(0.051 f_{\rho e},10^{-4}\mathrm{nT}^2/\mathrm{Hz})$; the spectrum amplitudes
 at $f_0$ are linearly interpolated from the two nearest points.
The results do not change if we choose another way to adjust the amplitudes in order to bring the spectra together. 
This rescaling allows us to fix the last free parameter in Eq.~(\ref{eq:model-1param}), the amplitude  to a value $A_0$, which 
is now related to $P_0$ at $f_0$. Thus, we can compare the shape of 3344 spectra with the function 
\begin{equation}\label{eq:3}
P_{\text{model}}(f/f_{\rho e}) = A_0(f/f_{\rho e})^{-8/3}\exp{(-1.8f/f_{\rho e})}. 
\end{equation}
This model passes through the data without any 
fitting; only the frequency is normalized to $f_{\rho e}$, and the amplitude is rescaled at the point $(f_0,P_0)$, see the dashed line in Figure~\ref{fig:spec3344}(c).
 Note that the  dispersion of the data points at the lowest and highest frequency ends can be due to the non simultaneous $T_e$ measurements. Moreover, the lowest frequency can be affected  as well by the proximity of the ion characteristic scales, and the highest frequencies can be affected by the SCM noise.

\section{Conclusion and discussion}

These results together with the previous observations at 1~AU \citep{alexandrova12}, indicate that
at kinetic scales smaller than the ion characteristic scales, the spectrum in the fast wind keeps its shape $\sim f^{-8/3} \exp{(-f/f_{d})}$
independently of the radial distance from the Sun, from 0.3 to 1~AU, with an exponential falloff, reminiscent of the dissipation range of the neutral fluid turbulence. 
The equivalent of the Kolmogorov scale $\ell_d$, where the dissipation of the electromagnetic cascade is expected to  take place, is controlled by the electron Larmor radius  $\rho_e$ for these radial distances. 
Precisely, here, with Helios we find $\ell_d\simeq 1.8 \rho_e$, and previously, with Cluster at 1~AU, we observed $\ell_d\simeq 1.4\rho_e$ \citep{alexandrova12}. The constant in front of $\rho_e$ seems to be weakly dependent on $R$. This will be verified in a future study with PSP and Solar Orbiter.

The equivalence between $\ell_d$ and $\rho_e$ is not a trivial result. First,  the electron Larmor radius is not the only characteristic length at such small scales. Closer to the Sun, the electron inertial length $\lambda_e$ becomes larger than the Larmor radius $\rho_e$, but as observed here, it is still with $\rho_e$ and not with $\lambda_e$  that the ``dissipation'' scale correlates.  
Second,  in neutral fluids, the dissipation scale $\ell_d$ 
is  much larger than the mean free path, so that the dissipation range is described within the fluid approximation.  In the solar wind between 0.3 and 1~AU, as we showed, $\ell_d$  is defined by  $\rho_e$ scale. 
In the vicinity of $\rho_e$ the protons are completely kinetic, and electrons start to be kinetic.  
Third, it appears puzzling that the dissipation scale in space plasma is fixed to a given plasma scale. 
It is well known in neutral fluids that  the dissipation scale $\ell_d$ depends on the energy injection rate 
$\varepsilon$ and thus on the amplitude of  turbulent spectrum in the following way: 
 $A\sim \varepsilon^2/3 \sim \ell_d^{-8/3}$ \cite[e.g.,][]{frisch1995,alexandrova09}.
	Is $\rho_e$ independent of the energy injection?  
We  found previously that the turbulent spectrum amplitude is anticorrelated with $\rho_e$ \cite{alexandrova09}; that is, it seems that the electron Larmor radius is sensitive to the turbulence level and thus to the energy injection.  
	We expect to verify this point with PSP and Solar Orbiter  data in future studies.

The results presented here may suggest that around the $\rho_e$ scale the electron Landau damping is at work to dissipate magnetic fluctuations into electron heating: this is found in 3D gyrokinetic simulations \citep{TenBarge2013} and  in analytical models of strong kinetic Alfv\'en wave (KAW) turbulence \citep{PassotSulem2015ApJ,Schreiner2017} and can be explained by the weakened cascade model of \citet{howes11pop}.  
However, in these theoretical and numerical works, the particle distributions are assumed to be Maxwellian, which is not the case in solar wind.

	It seems that the electron Landau damping is not the only possible dissipation mechanism. \citet{Parashar2018ApJL} observed that the spectral curvature at electron scales is sensitive to the $\rho_e$ scale (i.e., to $\beta_e$) in 2D 
	Particle-in-cell simulations, where the direction parallel to $\bm{B}_0$ is not resolved, so that the Landau damping cannot be effective. 
	 \citet{rudakov11} studied the weak KAW turbulence and showed that a non-Maxwellian electron distribution function  has a significant effect on the cascade:  the linear Landau damping leads to the formation of a plateau in the parallel electron distribution function $f(V_{e\|})$, for $V_A<V_{e\|}<V_{e,th}$, which reduces the Landau damping rate significantly. These authors studied the nonlinear scattering of waves by plasma particles and concluded that, for the solar wind parameters, this	scattering is the dominant process at kinetic scales, with the dissipation starting at the $\lambda_e$ scale. To date, we have not measured in the solar wind a plateau in $f(V_{e\|})$ between the Alfv\'en speed $V_A$ and the electron thermal speed $V_{e,th}$. 
	 Such a distribution may exist, but would be very difficult to observe because of instrumental effects such as the spacecraft potential and photoelectrons.
However, it is not clear  to what extent the quasi-linear results based on the Landau damping or the weakly non-linear model of \citet{rudakov11}
	are relevant when non-linear coherent structures \citep{greco16,perri12a}
importantly contribute to the turbulent power spectrum on kinetic scales.

\begin{figure}
	\includegraphics[width=9.0cm]{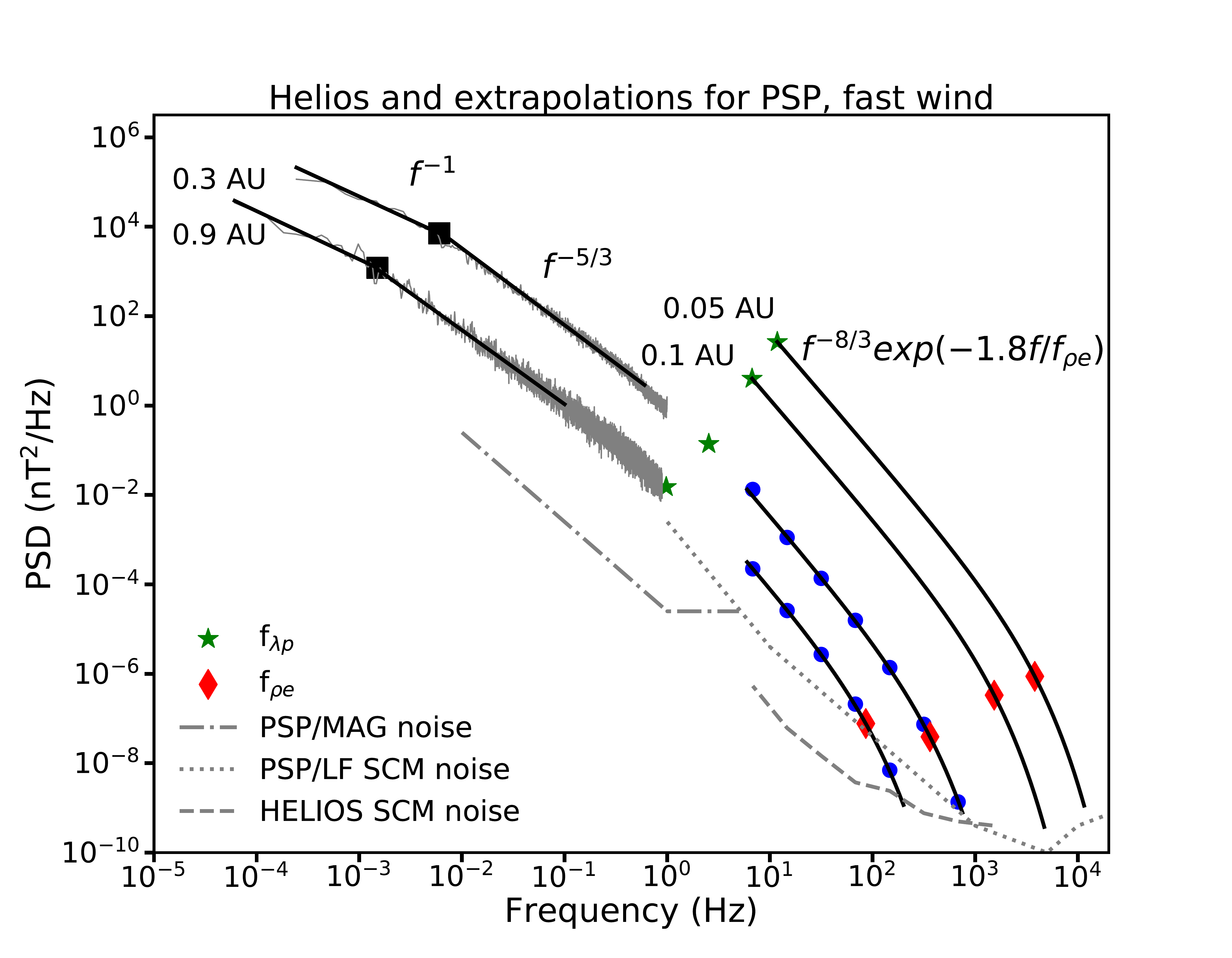}
	\caption{The complete turbulent spectrum from energy injection scales  up to the sub-electron scales at 0.3 and 0.9~AU as measured by Helios. 
		The energy containing scales (which correspond to $\sim f^{-1}$ spectrum) and the MHD inertial range ($\sim f^{-5/3}$) are covered by the Helios--MAG instrument (gray lines). 
		The Helios--SCM instrument covers the kinetic scales (blue dots), studied in the present paper.
		The black solid lines indicate model functions  $f^{-1}$,  $f^{-5/3}$ and   $f^{-8/3}\exp{(-1.8f/f_{\rho_e})}$ at different frequency ranges. The two most energetic spectra at high frequencies are the extrapolations of the kinetic spectrum in the fast wind that we expect to measure with PSP at 0.05 and 0.1~AU. The dashed line gives Helios-SCM noise, the dashed-dotted and dotted lines indicate noise levels of the different magnetic sensors on PSP.
		The Doppler shifted ion inertial length $\lambda_p$ (green stars) marks the transition from the inertial to the kinetic range;  the electron Larmor radius $\rho_e$ (red diamonds) marks the dissipation  cutoff.}
	\label{fig:psp}
\end{figure}

\begin{table}
	\centering
	\caption{Mean plasma parameters at 4 radial distances from the Sun, corresponding to the spectra in Figure~\ref{fig:psp}.}
	\begin{tabular}{ c | c c c c c c c c c }
		\hline
		$R$ (AU) & 0.9 & 0.3  & 0.1 & 0.05   \\ \hline
		
		$B_0$ (nT) & $7\pm2$& $41\pm 3$ & 280 & 990  \\ 
		$V$ (km/s)  & $705\pm35$ & $650\pm 40$ & 510 & 410 \\ 
		
		$n_p$ (cm$^{-3}$) & $4\pm1$ & $31\pm 4$ & 350 &1700 \\ 
		
		$T_p$ (eV) &$21\pm 5$ &$50\pm9$ & 120 & 230 \\ 
		$T_e$ (eV) &$9 \pm 2$ &$15\pm 2$ & 19 & 25 \\ 
		$T_{p \perp}$ (eV) &$24\pm 5$ &$65\pm10$ & - & - \\ 
		$T_{e \perp}$ (eV) &$7\pm1$ &$12\pm1$ & - & -  \\ 
		$\beta_p$ & $0.8\pm 0.2$ & $0.5\pm 0.1$ & 0.2 & 0.15 \\ 
		$\beta_e$  &$0.2\pm0.1$ &$0.10\pm0.02$ & 0.04 & 0.02 \\ 
		$\lambda_p$ (km) &$108 \pm 14$ &$39\pm 3$ & 12 & 6  \\ 
		$\rho_p$ (km) &$101 \pm 31$ &$28\pm 3$& 6 & 2 \\ 
		$\lambda_e$ (km)  &$2.5 \pm 0.3$ &$0.9 \pm 0.1$ & 0.3 & 0.1  \\ 
		$ \rho_e$ (km) &$1.3\pm 0.4$ &$0.3 \pm 0.02$ & 0.05 & 0.02  \\ 
		$f_{cp}$ (Hz) &$0.10 \pm0.03 $&$0.6 \pm 0.05$ & 4 & 15  \\ 
		$f_{\lambda p}$ (Hz)  &$1.0 \pm 0.1$& $2.6 \pm 0.3$ & 7 & 12 \\ 
		$f_{\rho p}$ (Hz) &$1.1 \pm 0.3$ &$3.6 \pm 0.5$ & 14 & 30  \\ 
		$f_{\lambda e}$ (Hz)  & $44\pm 6$&$110 \pm 10$ & 300 & 500  \\ 
		$f_{\rho e}$ (Hz) &$90 \pm 30$ &$360\pm 40$ & 1530 & 3800 \\ 
		$f_{ce}$ (Hz) &$200 \pm 60 $&$1150\pm 80$ & 7800 & 28000  \\ \hline
	\end{tabular}
	\label{tab:tab1}
\end{table}

Let us now put our observations in a more general context of the solar wind turbulence. 
Figure~\ref{fig:psp}  shows a complete turbulent spectrum covering the energy containing scales ($\sim f^{-1}$ spectral range), the inertial range at MHD scales ($\sim f^{-5/3}$ range), and the kinetic scales, as observed at 0.3 and 0.9~AU by Helios in the fast wind. The mean plasma parameters for the time intervals used here are given in Table~\ref{tab:tab1}.

We expect that the spectral properties we observe are generic for plasma turbulence at  sub-ion to electron scales. The two most energetic spectra at high frequencies in Figure~\ref{fig:psp} are the extrapolations of the
	kinetic spectrum that we expect to observe in the fast solar wind with PSP at 0.05 and 0.1~AU (see the Appendix for more details). 
Indeed, the beginning of this kinetic spectrum following an $f^{-8/3}$--law between $\sim 10$ and $100$~Hz was recently observed by PSP at 35.7 solar radii (0.166~AU) \cite{Bale2019Nat,Bowen2020PRL}. 
Future PSP observations closer to the Sun will show how the empirical picture of the kinetic turbulence given here may change.

\section*{Appendix: Extrapolation of turbulent spectra closer to the Sun}

To plot the extrapolations of the kinetic spectra at 0.05 and 0.1~AU 
	in Figure~\ref{fig:psp},  we assume that the turbulence level will increase together with the mean field, keeping $\delta B/B_0\sim \text{const}$, as observed in the solar wind, \cite[e.g.,][]{beinroth81,bourouaine12}.
In the inner heliosphere, where $\beta<1$, the end of the Kolmogorov scaling is expected to happen at the proton inertial length $\lambda_p$  \citep{bourouaine12,chen14} (see green stars). 
The exponential falloff at the end of the electromagnetic cascade is defined by the local $\rho_e$, as we confirm in this study. 
To determine the Doppler shifted frequencies where  $\lambda_p$ and $\rho_e$ will appear in the extrapolated spectra ($f_{\lambda p}=V/2\pi \lambda_p$   and $f_{\rho e}=V/2\pi \rho_e$), we use plasma parameters (proton density $n_p$, electron temperature $T_e$, magnetic field $B_0$, and solar wind speed $V$) extrapolated from the in-situ Helios measurements (from 0.3 to 0.9~AU). 
These latter extrapolations have been performed by connecting the gradient of the Helios density measurements to the one measured remotely from coronal white light eclipse observations. 
More precisely, we have retrieved the radial variations of both the electron density $n_e(R)$ (which we assume for simplicity to be equal to $n_p(R)$) and bulk speed $V(R)$ all the way down to the low corona by (i) imposing that the density matches both the 0.3 to 1~AU Helios density observations and the coronal density observations obtained remotely by \citet{Sittler1999ApJ} 
and (ii)  imposing the conservation of the mass flux $n_e(R) V(R) R^2 = \text{const}$.
The plasma parameters used for the extrapolated spectra as well as for the time intervals of the Helios measurements are summarized in Table~\ref{tab:tab1}.

\acknowledgments
O.A., V.K.J., and M.M. are supported by the French Centre National d’Etude Spatiales (CNES). P.H. acknowledges Grant No. 18-08861S from the Czech Science Foundation.
O.A. thanks F. Neubauer and L. Matteini for discussions and C. Lacombe for reading this manuscript. 

\paragraph*{Data} 
The Helios--1 data are available on the Helios data archive (http://helios-data.ssl.berkeley.edu/). 
\paragraph*{Software}  The routine used to fit the data with the model, Eq.(1), is optimize.curve$\_$fit from scipy$/$python \citep{2020SciPy-NMeth}. 
\paragraph*{Correspondence} Correspondence 
should be addressed to O. Alexandrova~(email: olga.alexandrova@obspm.fr).

\bibliography{helios-spectra-2020}

\end{document}